\documentclass[aoas,MSNbibl,nameyear,seceqn,dvips]{arximspdf}
\usepackage{dcolumn}
\usepackage{graphicx}

\doi{10.1214/14-AOAS764} 
\volume{8}
\issue{4}
\pubyear{2014}
\firstpage{2203}
\lastpage{2222}
\docsubty{FLA}

\makeatletter
\newcolumntype{d}[1]{D{.}{.}{#1}}

\renewcommand{\mid}{|}
\newtheorem{theorem}{Theorem}
\newtheorem{proposition}[theorem]{Proposition}
\newcommand{\mutAlpha}{{\alpha}}
\newcommand{\mutA}{{u_{01}}}
\newcommand{\mutBeta}{{\beta}}
\newcommand{\mutB}{{u_{10}}}
\newcommand{\meanFitness}{\bar\sigma}
\newcommand{\jacobi}{{R}}
\newcommand{\jacobiLength}{{c}}
\newcommand{\jacobiLengthMatrix}{\mathbf{C}}
\newcommand{\modJacobi}{{H}}
\newcommand{\eValNeutr}{{\lambda}}
\newcommand{\eValMat}{\bolds{\Lambda}}
\newcommand{\eValNeutrMat}{\bolds{\Lambda}}
\newcommand{\eVec}{\mathbf{w}}
\newcommand{\eVecMatrix}{\mathbf{W}}
\newcommand{\vecEFun}{\mathbf{B}}
\newcommand{\eFunLength}{{c}}
\newcommand{\eFunLengthMatrix}{\mathbf{D}}
\newcommand{\genFull}{{\mathcal{L}}}
\newcommand{\alleleFreqRV}{\mathbf{Y}}
\newcommand{\hilbert}{{L}}
\newcommand{\numSamples}{{n}}
\newcommand{\numPoints}{{K}}
\newcommand{\diffTime}{{\tau}}
\newcommand{\initialDist}{{\rho}}
\newcommand{\vecAuxiliary}{\mathbf{a}}
\newcommand{\vecForward}{\mathbf{b}}
\newcommand{\coeffForward}{{b}}
\newcommand{\bOne}{\mathbf{1}}
\newcommand{\diag}{\operatorname{diag}}
\newcommand{\N}{{\mathbb{N}}}
\renewcommand{\L}{{\mathcal{L}}}
\makeatother

\begin{document}
\begin{frontmatter}

\title{A novel spectral method for inferring general diploid selection
from time series genetic data}
\runtitle{A spectral method for inferring selection}

\begin{aug}
\author[A]{\fnms{Matthias} \snm{Steinr\"{u}cken}\thanksref{T2,T3}\ead[label=e1]{steinrue@stat.berkeley.edu}},
\author[B]{\fnms{Anand} \snm{Bhaskar}\thanksref{T2}\ead[label=e2]{bhaskar@eecs.berkeley.edu}}
\and
\author[C]{\fnms{Yun~S.} \snm{Song}\corref{}\thanksref{T2}\ead[label=e3]{yss@stat.berkeley.edu}}
\runauthor{M. Steinr\"{u}cken, A. Bhaskar and Y.~S. Song}
\affiliation{University of California, Berkeley}
\address[A]{M. Steinr\"{u}cken\\
Department of Statistics\\
\quad and Computer Science Division\\
University of California, Berkeley\\
Berkeley, California 94720\\
USA\\
\printead{e1}}
\address[B]{A. Bhaskar\\
Computer Science Division\\
University of California, Berkeley\\
Berkeley, California 94720\\
USA\\
\printead{e2}}
\address[C]{Y.~S. Song\\
Department of Statistics\\
\quad and Computer Science Division\\
and\\
Department of Integrative Biology\\
University of California, Berkeley\\
Berkeley, California 94720\\
USA\\
\printead{e3}}
\end{aug}
\thankstext{T2}{Supported in part by NIH Grant R01-GM094402 and a
Packard Fellowship for Science and Engineering.}
\thankstext{T3}{Supported in part by DFG research fellowship STE 2011/1-1.}

\received{\smonth{10} \syear{2013}}
\revised{\smonth{3} \syear{2014}}

%
\begin{abstract}
The increased availability of time series genetic variation data from
experimental evolution
studies and ancient DNA samples has created new opportunities to
identify genomic regions under selective pressure and to \mbox{estimate} their
associated fitness parameters.
However, it is a challenging problem to compute the likelihood of
nonneutral models for the
population allele frequency dynamics, given the observed temporal DNA
data. Here, we develop a novel spectral algorithm to analytically and
efficiently integrate over all possible frequency trajectories between
consecutive time points. This advance circumvents the limitations of
existing methods which require fine-tuning the discretization of the
population allele frequency space when numerically approximating
requisite integrals. Furthermore, our method is \mbox{flexible} enough to
handle general diploid models of selection where the heterozygote and
homozygote fitness parameters can take any values, while previous
methods focused on only a few restricted models of selection. We
demonstrate the utility of our method on simulated data and also apply
it to analyze ancient DNA data from genetic loci associated with coat
coloration in horses. In contrast to previous studies, our exploration
of the full fitness parameter space reveals that a
heterozygote advantage form of balancing selection may have been acting
on these loci.
\end{abstract}

%
\end{frontmatter}

\setcounter{footnote}{2}

\section{Introduction}\label{secintroduction}\label{sec1}
Natural selection is a fundamental evolutionary process and finding
genomic regions experiencing selective pressure has important
applications, including identifying the genetic basis of diseases and
understanding the molecular basis of adaptation.
There has been a long line of theoretical and experimental research
devoted to modeling and detecting selection acting at a given locus.
Several earlier works have considered modeling the stationary
distribution of allele frequencies in a population undergoing
nonneutral evolution
[\citeauthor{Fearnhead2003} (\citeyear{Fearnhead2003,Fearnhead2006}), \citet{Genz2003,Stephens2003}].
More recently, there has been growing interest to utilize time series
genetic variation data to enhance our ability to infer allele frequency
trajectories, thereby enabling better estimates of selection
parameters. For example, the sequencing of samples over several
generations in experimental evolution of a population (e.g., Bacteria
[\citet{Wiser2013}], yeast [\citet{Lang2013}] and \emph
{Drosophila} [\citet{Burke2010,Orozco-terWengel2012}]) under
controlled laboratory environments, or direct measurements in fast
evolving populations such as HIV [\citet{Shankarappa1999}], has
allowed us to better understand the genetic basis of adaptation to
changes in the environment.
Also, recent technological advances have given us the unprecedented
ability to acquire ancient DNA samples (e.g., for humans [\citet
{Hummel2005}], ancient hominids [\citet{Green2010,Reich2010}]
and horses [\citet{Ludwig2009,Orlando2013}]), providing useful
information about allele frequency trajectories over long evolutionary
timescales.

Most methods for analyzing times series DNA data model the underlying
population-wide allele frequency as an unobserved latent variable in a
hidden Markov model (HMM) framework, in which the sample of alleles
drawn from the population at a given time is treated as a noisy
observation of the hidden population allele frequency. In this
framework, computing the probability of observing time series genetic
variation data involves integrating over all possible hidden
trajectories of the population allele frequency. For short evolutionary
timescales, a discrete-time Wright--Fisher model of random mating is
often used to describe the dynamics of the population allele frequency
in the underlying HMM. This approach has been used to estimate the
effective population size from temporal allele frequency variation,
assuming a neutral model of evolution [\citet{Williamson1999}].
More recently, temporal and spatial variations of advantageous alleles
have been investigated through an HMM framework that can incorporate
migration between multiple subpopulations [\citet{Mathieson2013}].

If the evolutionary timescale between consecutive sampling times is
large, it can become computationally cumbersome to work with
discrete-time models of reproduction. However, by a suitable rescaling
of time, population size and population genetic parameters, one can
obtain a continuous-time process (the Wright--Fisher diffusion) which
accurately approximates the population allele frequency of the
discrete-time Wright--Fisher model. The key quantity needed when
applying the diffusion process is the transition density function,
which describes the probability density of the allele frequency
changing from value $x$ to value $y$ in time $t$. This transition
density function satisfies a certain partial differential equation
(PDE) with coefficients that depend on the mutation and selection
parameters. \citet{Bollback2008} have used a finite-difference
numerical method to approximate the solution to the PDE and
incorporated the results into the aforementioned HMM framework to infer
the strength of selection from time series data.
Recently, an alternative approach [\citet{Malaspinas2012}] based
on a one-step Markov process has been proposed to compute the necessary
transition densities.
In both of these approaches, the allele frequency space has to be
discretized finely enough in order to reliably approximate various
numerical integrals that are needed for computing the HMM likelihood.
The efficiency and accuracy of these grid-based numerical methods
depend critically on the spacing and distribution of the discrete grid
points. Furthermore, an appropriate choice of this discretization
scheme could be strongly dependent on the underlying population genetic
parameters.
Another limitation of these previous works is that only a few
restricted models of selection have been considered.
\citet{Feder2014} recently developed a likelihood-ratio test for
identifying signatures of selection from time series data in which they
combined a deterministic model and a Gaussian noise process. This
approximation is less accurate than the diffusion approximation, but it
facilitates computation and seems sufficiently accurate provided that
the allele frequency does not get too close to the boundaries during
the period of observation.

In this paper, we develop a novel algorithm based on the spectral
method to circumvent the limitations mentioned above.
Specifically, instead of approximating the solution to the PDE
numerically, we utilize a method recently developed by \citet
{Song2012} which finds an explicit spectral representation of the
transition density as a function of $x$, $y$ and $t$. We show that the
probability of observing a given time series data set can be computed
analytically by combining the spectral representation with the forward
algorithm for HMMs to \emph{efficiently} and \emph{analytically}
integrate over all population allele frequency trajectories.
The key idea in our work is to represent the intermediate densities in
the forward algorithm in the basis of eigenfunctions of the
infinitesimal generator of the Wright--Fisher diffusion process.
Exploiting the spectral representation of the transition density, we
can then efficiently compute the coefficients in this basis
representation. Furthermore, since this spectral representation applies
to general diploid models of selection, we are able to leverage this
representation to consider more complex models of selection than
previously possible. We first demonstrate the accuracy of our method on
simulated data. We then apply the method to analyze time series ancient
DNA data from genetic loci (ASIP and MC1R) that are associated with
horse coat coloration. In contrast to the conclusions of previous
studies which considered only a few special models of selection
[\citet{Ludwig2009,Malaspinas2012}], our exploration of the full
parameter space of general diploid selection reveals that a
heterozygote advantage form of balancing selection may have been acting
on these loci. We implemented the algorithms described in this paper in
a publicly available software package called \emph
{spectralHMM}.\footnote{Available from \url{http://spectralhmm.sf.net}.}

The remainder of this paper is organized as follows. In Section~\ref
{secmodel} we formally introduce the HMM framework and describe the
details of our spectral algorithm. The proofs of the theoretical
results underlying our algorithm are provided in the supplemental
article [\citet{sbssupp2014}]. In Section~\ref{secresults} we use
simulated data to investigate the statistical properties of our maximum
likelihood estimator and also apply our method to analyze the
aforementioned ancient DNA data for the loci associated with horse coat
coloration [\citet{Ludwig2009}]. We conclude in Section~\ref
{secdiscussion} with a discussion of future extensions of our model.

\section{Method}
\label{secmodel}

Here we provide a formal description of the time series data considered
in this paper and
present our inference method for analyzing such data.

\subsection{Time series allele frequency data}
\label{secdata}

\begin{figure}

\includegraphics{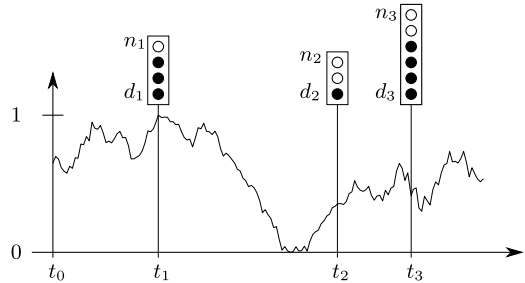}

\caption{In this example, samples of size $\numSamples_1 = 4$,
$\numSamples_2 = 3$ and $\numSamples_3 = 6$ (illustrated by the total
number of circles) are taken at times ${t}_1$, ${t}_2$
and ${t}_3$, respectively. The observed number of derived alleles
(filled circles) is $d_1 = 3$, $d_2
= 1$ and $d_3 = 4$. The initial time is ${t}
_0$, and the curve indicates a particular trajectory of the underlying
population allele frequency $\alleleFreqRV({t}) \in[0,1]$.}\label{figdata}
\end{figure}

The data we analyze consist of genotype samples obtained from
individuals at $K$ distinct times ${t}_1 < \cdots< {t}
_\numPoints$ in the past (given in years). The present time is denoted by
${t}_{\mathrm{present}} \geq t_\numPoints$. At each time point
${t}_{k}$, a sample of $\numSamples_{k}\in\N$
individuals is randomly drawn from the population. We assume that the
locus under consideration is biallelic, and that the identities of the
ancestral allele $A_0$ and the derived allele
$A_1$ are known. We also assume that the allele
$A_1$ became selected at some time ${t}_0 \leq
{t}_1$. We use $d_{k}$ to denote the
number of derived alleles in the sample of $\numSamples_{k}$
alleles drawn at time ${t}_{k}$, where $0 \leq
d_{k}\leq\numSamples_{k}$. For
notational convenience, we use ${o}_{k}$ to denote
the tuple $({t}_{k}, \numSamples_{k},
d_{k})$ and ${O}_{[i\dvtx j]}$ to
denote the partial sequence of observations ${o}_i,
{o}_{i+1}, \ldots, {o}_j$. Figure~\ref{figdata} shows
an example of a time series allele frequency data set with samples
drawn at three time points.

%

\subsection{The diffusion approximation}
\label{secdiffusionmodel}

Consider a locus evolving according to a discrete
Wright--Fisher model of random mating with an effective population size
of $N_e$ diploids. Let $\mutA$ be the per generation probability
of mutation from the ancestral allele $A_0$ to the derived
allele $A_1$, and $\mutB$ the probability of the
reverse mutation. We use ${s}_i$ to denote the selection coefficient
of an individual with $i$ copies of the derived allele
$A_1$, where $0 \leq i \leq2$. Without loss of
generality, we can assume that ${s}_0 = 0$.
In each generation of reproduction, an offspring randomly chooses a
parent having $i$ copies of the derived allele with probability
proportional to $1 + {s}_i$.

Consider the scaling limit where the population size $N_e\to
\infty$ while the unit of time is rescaled by $N_e$ and the
population-scaled parameters ($2N_e{{s}_{1}}$, $2N_e
{{s}_{2}}$, $4N_e\mutA$, $4N_e\mutB$) approach some
constants. In this limit, the trajectory of the population frequency of
allele $A_1$ follows a Wright--Fisher diffusion process
[\citet{Ewens2004}]. The unit of time $\diffTime$ in this
diffusion approximation is related to the physical unit of time
${t}$ as
\[
\diffTime= {t}/(2N_e{g}),
\]
where ${g}$ is the average number of years per generation of
reproduction. Similarly, we let $\diffTime_{k}$ denote the
population-scaled versions of the physical times ${t}_{k}
$, where
%
\begin{eqnarray}
\diffTime_{k}&=& {t}_{k}/(2N_e {g}).
\label{eqdiffTimepointsIdxdef}
\end{eqnarray}
The population-scaled selection and mutation parameters of the
Wright--Fisher diffusion process are related to the corresponding
parameters in physical units as
%
\begin{eqnarray}
{\sigma}_i &=& 2N_e{s}_i, \label{eqselSigmadef}
\\
\mutAlpha&=& 4N_e\mutA, \label{eqmutAlphadef}
\\
\mutBeta&=& 4N_e\mutB. \label{eqmutBetadef}
\end{eqnarray}
From here on, we use the above population-scaled parameters when
describing our analysis of the Wright--Fisher diffusion.
The initial population frequency of the allele $A_1$
when it became selected at time $\diffTime_0$ is distributed according
to the density function $\initialDist({y})$.
In this paper, we are interested in estimating the selection
coefficients of the heterozygote and $A_1$-homozygote
(${{s}_{1}}$ and ${{s}_{2}}$, resp.) given the
other population genetic parameters and
assuming that the allele $A_1$ became selected at time
$\diffTime_0$.

\subsection{Hidden Markov model framework}
\label{sechmm}

To analyze the time series data described earlier, we employ a hidden
Markov model (HMM) framework as in \citet{Bollback2008}.
In this approach, the population-wide frequency $\alleleFreqRV
(\diffTime)$ of the $A_1$ allele at time $\diffTime$
is modeled as an unobserved hidden variable (see Figure~\ref{figdata}). We
denote a realization of the frequencies at the sampling times
$\diffTime_{k}$ by ${y}_{k}\equiv
\alleleFreqRV(\diffTime_{k})$. The initial frequency at time
$\diffTime_0$ is distributed according to the density function
$\initialDist$, that is, $\alleleFreqRV(\diffTime_0) \sim
\initialDist$. For example, the density function $\initialDist
({y}) = \delta({y}-1/(2N_e))$ models the case
where the selected allele $A_1$ arose as a \emph{de
novo} mutation in one individual of the population at time $\diffTime_0$.

The probability of transitioning from frequency ${y}
_{{k}-1}$ at time ${t}_{{k}-1}$ to frequency
${y}_{{k}}$ at time ${t}_{{k}}$ is
described by the transition density function ${p}_\Theta(\diffTime
_{{k}} - \diffTime_{{k}-1}; {y}_{{k}
-1}, {y}_{{k}})$ of the Wright--Fisher diffusion process,
where $\Theta= ({{\sigma}_{1}}, {{\sigma}_{2}}, \mutAlpha, \mutBeta,
\diffTime
_0, N_e)$ and $\diffTime_{k}$ are population-scaled
parameters as given in equations (\ref
{eqdiffTimepointsIdxdef})--(\ref{eqmutBetadef}).
The observations in the HMM are the number of copies
$d_{k}$ of the allele $A_1$
among the $\numSamples_{k}$ alleles in the sample drawn at
time ${t}_{k}$. The probability of such an observation
at time~${t}_{k}$ with population allele frequency
${y}_{k}$ is given by the probability mass function
${\xi}(d_{k}; \numSamples_{k},
{y}_{k})$ of a binomial distribution
\[
{\xi}({d}_{k}; \numSamples_{k}, {y}_{k}):=\pmatrix{
\numSamples_{k}\cr{d}_{k}} {y}_{k}
^{{d}_{k}} (1 - {y}_{k} )^{\numSamples_{k}- {d}_{k}}.
\]

To compute the probability ${\mathbb{P}}_\Theta\{ {O}
_{[1\dvtx\numPoints]} \}$ of observing the data ${O}
_{[1\dvtx\numPoints]}$ under the model parameters $\Theta$, we introduce
the forward density functions $f_{k}$, given by
%
\begin{equation}
f_{k}({y}) \,d{y}:={\mathbb{P}} _\Theta\bigl\{
{O}_{[1\dvtx{k}]}, \alleleFreqRV(\diffTime_{k}) \in d{y} \bigr\},
\qquad{k}\in\{0,1,\ldots,\numPoints\}. \label{eqforwardDensity}
\end{equation}
The function $f_{k}$ is the joint density of the
observed data up to time $\diffTime_{k}$ and the hidden
population allele frequency at time $\diffTime_{k}$. We also
find it convenient to consider a second auxiliary density function,
$g_{k}$, given by
%
\begin{equation}
g_{k}({y}) \,d{y}:={\mathbb{P}} _\Theta\bigl\{
{O}_{[1\dvtx{k}-1]}, \alleleFreqRV(\diffTime_{k}) \in d{y} \bigr\},
\qquad{k} \in\{1,\ldots,\numPoints\}. \label{eqauxiliaryDensity}
\end{equation}
This function $g_{k}$ is the joint density of
the observed data up to time $\diffTime_{{k}- 1}$ and the
hidden frequency at $\diffTime_{k}$.
The forward density function $f_0$ is given by the
density function for the initial allele frequency as
\[
f_0 ({y}) = \initialDist({y}).
\]
Since we approximate the time evolution of the hidden population allele
frequency by the Wright--Fisher diffusion, we can get a recurrence
relation between the density functions $g_{k}$
and $f_{{k}-1}$ by integrating over all possible
allele frequencies at~$\diffTime_{{k}- 1}$:
%
\begin{equation}
\label{eqtransitionintegral} g_{k}({y}) = \int_0^1
f_{{k}-1} ({x}) {p}_\Theta(\diffTime_{k}-
\diffTime_{{k}-1};{x},{y}) \,d{x},
\end{equation}
where ${k}\in\{ 1,\ldots,\numPoints\}$.
Using the binomial distribution for sampling ${d}
_{k}$ derived alleles out of $\numSamples_{k}$
individuals at time $\diffTime_{k}$, we get another recurrence
relation between the density functions $f_{k}$
and $g_{k}$ as follows:
%
\begin{equation}
\label{eqbinomialemission} f_{k}({y}) = g _{k}({y}) {
\xi}({d}_{k}; \numSamples_{k}, {y}).
\end{equation}
Finally, the probability ${\mathbb{P}}_\Theta\{ {O}
_{[1\dvtx\numPoints]} \}$ of observing the data is computed by integrating
over all possible hidden frequencies at the last sampling time:
%
\begin{equation}
\label{eqfinalintegral} {\mathbb{P}}_\Theta\{ {O}_{[1\dvtx\numPoints]}
\} = \int
_0^1 f_\numPoints({y}) \,d{y}.
\end{equation}
Note that the equations above describe a forward-in-time procedure for
computing the probability of the data ${O}
_{[1\dvtx\numPoints]}$, where the intermediate density functions have a
natural interpretation.

While (\ref{eqtransitionintegral}), (\ref{eqbinomialemission})
and (\ref{eqfinalintegral}) succinctly describe the sampling
probability of the data ${O}_{[1\dvtx\numPoints]}$, no
analytic solutions to the integrals in (\ref{eqtransitionintegral})
and (\ref{eqfinalintegral}) are known. In the previous approaches
mentioned in the
\hyperref[sec1]{Introduction}, these integrals were approximated numerically by
discretizing the allele frequency state space. The accuracy of these
approximations depends critically on the careful choice of the
discretization grid. We present an analytical solution to this problem
which obviates the need for such a discretization.


\subsection{Spectral representation of the transition density}
\label{secspectralrepresentation}

The biallelic\break Wright--Fisher diffusion with general diploid selection
has the infinitesimal generator $\genFull$ given by
%
\begin{eqnarray}
\label{eqgenFull} \genFull&=& {\L_0}+ 2 {x}(1-{x} ) \bigl[ {{
\sigma}_{1}}(1 - 2{x}) + {{\sigma}_{2}} {x} \bigr]
\frac{\partial}{\partial
{x}},
\end{eqnarray}
where ${\L_0}$ is the infinitesimal generator of the diffusion
process without selection, given by
%
\begin{eqnarray}
\label{eqgenNeutral} {\L_0}&=& \frac{1}{2}{x}(1-{x})
\frac{\partial^2}{\partial{x}^2} + \frac{1}{2} \bigl[\mutAlpha(1-{x}) -
\mutBeta{x} \bigr]
\frac{\partial}{\partial{x}}.
\end{eqnarray}
We refer the reader to \citet{Ewens2004} for more details about
the Wright--Fisher diffusion.
\citet{Song2012} developed an efficient method to compute the
eigenvalues and eigenfunctions of $\genFull$, and we utilize that
method here. A brief summary of their approach is provided below.

To approximate the spectral decomposition of the operator $\genFull$,
consider the functions
%
\begin{eqnarray}
\label{eqmodJacobi} \modJacobi_{m}^{(\Theta)}({x})&:=& e^{-\meanFitness({x})/2}
\jacobi_{m}^{(\mutAlpha,\mutBeta)}(x),
\end{eqnarray}
where $\meanFitness({x}):=4 {{\sigma}_{1}}
{x}(1-{x}) + 2 {{\sigma}_{2}}
{x}^2$ is the mean fitness of the population and
$\jacobi_{m}^{(\mutAlpha,\mutBeta)}({x})$ are a
rescaled version of the classical orthogonal Jacobi polynomials and are
defined in Section~B of the supplemental article
[\citet{sbssupp2014}]. The $\alpha$ and $\beta$ parameters in
(\ref{eqmodJacobi}) are the\break population-scaled mutation rates given in
(\ref{eqmutAlphadef}) and (\ref{eqmutBetadef}). The set\break $ \{
\modJacobi_{m}^{(\Theta)}({x}) \}_{{m}
\in\N_0}$ forms a basis for the Hilbert space $\hilbert^2
([0,1],{\pi} )$ of real-valued functions on $[0,1]$ that are
square integrable with respect to the stationary density ${\pi}$ of
the diffusion generator $\genFull$. Specifically,
%
\begin{equation}
\label{eqstationarymeasure} {\pi}({x}) = e^{\meanFitness({x})}
{x}^{\mutAlpha-1}(1-{x})^{\mutBeta-1}.
\end{equation}
The basis elements $\modJacobi_{m}^{(\Theta
)}({x})$ are orthogonal with respect to the inner
product ${\langle\cdot, \cdot\rangle}_{{\pi}}$ defined by
${\langle f, g \rangle}_{{\pi}} = \int_0^1 f(x) g(x) {\pi}(x) \,dx$.

In the basis $ \{ \modJacobi_{m}^{(\Theta
)}({x}) \}_{{m}\in\N_0}$, the operator
$\genFull$ is given by the matrix
%
\begin{equation}
\label{eqoperatormatrix} \mathbf{M}:=- \Biggl( \eValNeutrMat^{(\mutAlpha
,\mutBeta)} + \sum
_{l=0}^4 {q}^{(\Theta)}_l
\mathbf{G}^l \Biggr),
\end{equation}
where $\eValNeutrMat^{(\mutAlpha,\mutBeta)}:=\diag(
\eValNeutr^{(\mutAlpha,\mutBeta)}_0, \eValNeutr^{(\mutAlpha,\mutBeta
)}_1, \ldots)$ is a diagonal matrix containing the
eigenvalues of the neutral diffusion generator ${\L_0}$,
$\mathbf{G}:= ({G}_{{n},{m}}^{(\mutAlpha,\mutBeta)} )_{{n},{m}\in\N_0}$
is the matrix of
coefficients from the three-term recurrence relation for the Jacobi
polynomials $\jacobi_{m}^{(\mutAlpha,\mutBeta
)}({x})$, and ${q}^{(\Theta)}_l$ are constant
coefficients defined in Section~C of the supplemental
article [\citet{sbssupp2014}].
Explicit expressions for the entries of $\eValNeutrMat^{(\mutAlpha
,\mutBeta)}$ and $\mathbf{G}$ are provided in equations (B.3)~and~(B.5), respectively, in
Section~B of the supplemental article.

The eigenvalues ${\lambda}_{n}$ of the full diffusion generator
$\genFull$ are given by the eigenvalues of $\mathbf{M}$, and the
coefficients of the eigenfunctions of $\genFull$ in the basis $ \{
\modJacobi_{m}^{(\Theta)}({x}) \}_{{m}
\in\N_0}$\vadjust{\goodbreak} are given by the eigenvectors of $\mathbf{M}$. In
particular, the eigenfunction ${B}_{n}$ of $\genFull$ is given by
%
\begin{equation}
\label{eqdefefun} {B}_{n}({x}) = \sum_{{m}=0}^\infty{w}
_{{n},{m}} \modJacobi_{m}^{(\Theta)} ({x}),
\end{equation}
where $\eVec_{n}= ({w}_{{n},0},{w}_{{n},1},\ldots)$ is the eigenvector
of $\mathbf{M}$ corresponding to eigenvalue~${\lambda}_{n}$.
We use $\eValMat= \diag( {\lambda}_0, {\lambda}_1, \ldots)$
to denote the diagonal matrix of eigenvalues of $\mathbf{M}$, and
$\eVecMatrix$ to denote the matrix with rows given by the eigenvectors
$\eVec_{n}$. As can be seen from (\ref{eqdefefun}), $\eVecMatrix
$ is the change-of-basis\vspace*{1pt} matrix between the basis of eigenfunctions
${B}_{n}$ of $\genFull$ and the basis $ \{ \modJacobi
_{m}^{(\Theta)}({x}) \}_{{m}\in\N_0}$.

The leading eigenvalues and the associated eigenvectors of the infinite
matrix~$\mathbf{M}$ can be approximated by the eigenvalues and
eigenvectors of sufficiently large submatrices of $\mathbf{M}$.
We refer the reader to \citet{Song2012} for a more detailed
empirical discussion on how the approximation accuracy varies for
different submatrix sizes and different parameter regimes. The
transition density function ${p}_\Theta(\diffTime; x, y)$ for the
probability density of the allele changing frequency from $x$ to $y$ in
time $\diffTime$ is given by the following spectral decomposition:
%
\begin{eqnarray}
\label{eqtdfspectraldecomposition} {p}_\Theta(\diffTime; {x}, {y}) &=&
\sum
_{{n}= 0}^{\infty} e^{-{\lambda}_{n}\diffTime} {\pi}(y)
\frac
{{B}_{n}(x) {B}_{n}(y)}{ {\langle{B}_{n}, {B} _{n} \rangle}_{{\pi}}}.
\end{eqnarray}

\subsection{Incorporating the spectral representation into the HMM}

Using the spectral decomposition of the transition density function in
(\ref{eqtdfspectraldecomposition}), we devise a dynamic programming
algorithm to compute the likelihood ${\mathbb{P}}_{\Theta}\{{O}
_{[1\dvtx\numPoints]}\}$. This algorithm recursively computes the density
functions $f_{k}$ and $g
_{k}$ given in (\ref{eqforwardDensity}) and (\ref
{eqauxiliaryDensity}), respectively.
To update these density functions efficiently, we represent them in the
basis of scaled eigenfunctions $ \{ {\pi}({y}) {B}
_{n}({y}) \}_{{n}\in\N_0}$ of the diffusion
generator $\genFull$.
More precisely, we express $f_{k}$ and
$g_{k}$ as
%
\begin{eqnarray}
f_{k}({y}) &=& {\pi}({y}) \vecForward_{k}\vecEFun({y}) =
\sum_{{n}=
0}^{\infty} \coeffForward_{{k},{n}}
{\pi}({y}) {B}_{{n}}({y}), \label
{eqforwardDensityrepresentation}
\\
g_{k}({y}) &=& {\pi}({y}) \vecAuxiliary_{k}\vecEFun({y}) =
\sum_{{n}=
0}^{\infty} {a}_{{k},{n}} {\pi}({y})
{B}_{{n}}({y}), \label{eqauxiliaryDensityrepresentation}
\end{eqnarray}
where we employ the vector notation
%
\begin{eqnarray}
\vecForward_{k}&:=&(\coeffForward_{{k},0},
\coeffForward_{{k},1}, \ldots), \label{eqvecForward}
\\
\vecAuxiliary_{k}&:=&({a}_{{k},0}, {a}_{{k},1}, \ldots),
\label{eqvecAuxiliary}
\\
\vecEFun({y}) &:=& \bigl({B}_0({y}), {B} _1({y}), \ldots
\bigr)^T. \label{eqvecEFun}
\end{eqnarray}

We now describe how the coefficient vectors $\vecAuxiliary_{k},
\vecForward_{k}$ and the probability ${\mathbb{P}}_\Theta\{
{O}_{[1\dvtx\numPoints]}\}$ can be computed efficiently.
All proofs can be found in Section~A of the supplemental
article [\citet{sbssupp2014}]. First, the\vadjust{\goodbreak} following proposition
determines the vector $\vecForward_0$ of coefficients for the initial
forward density function $f_0$:

\begin{proposition}\label{propinitial}
If the allele frequency at $\diffTime_0$ is distributed according to
the density function $\initialDist({y}) = \delta({y}
-{x})$, then the initial forward density function
$f_0$ in the basis $ \{ {\pi}({y}) {B}
_{n}({y}) \}_{{n}\in\N_0}$ has the vector of coefficients
\[
\vecForward_0 = \biggl( \frac{{B}_0({x}
)}{\eFunLength_0}, \frac{{B}_1({x})}{\eFunLength
_1},\ldots\biggr),
\]
where ${B}_{n}({x})$ is given by (\ref{eqdefefun}), and
$\eFunLength_{n}$
are
the squared norms of ${B}_{n}$
given by
%
\begin{equation}
\eFunLength_{n}= {\langle{B}_{n}, {B}_{n}
\rangle}_{{\pi}} = \sum_{{m}=0}^\infty
({w}_{{n},{m}} )^2 \jacobiLength_{n}^{(\mutAlpha,\mutBeta)},
\label{eqeFunLength}
\end{equation}
where $\jacobiLength_{n}^{(\mutAlpha,\mutBeta)}$ denote the
squared norms of the Jacobi polynomials
given in equation~\textup{(B.2)} in Section~\textup{B} of the supplemental article [\citet{sbssupp2014}].
\end{proposition}

In the case where the selected allele $A_1$ arises from
\emph{de novo} mutation at ${t}_0$ in one of the individuals in
the population, we set ${x}= 1/(2N_e)$ in
Proposition~\ref{propinitial}. We note that our framework allows us
to easily
model other distributions for the frequency of the mutant allele
$A_1$ when it became selected. For example, the initial
distribution of mutation-drift balance can be used to model selection
arising from standing genetic variation. Some of these initial
distributions are described in Section~D of the
supplemental article [\citet{sbssupp2014}].

The following theorem establishes how the representations of the
densities $f_{k}$ and $g
_{k}$, for ${k}> 0$, can be computed algebraically in a
recursive fashion:

%
\begin{theorem}\label{thmtransitionemission}
Let $\jacobiLengthMatrix:=\diag(\jacobiLength
_0^{(\mutAlpha,\mutBeta)}, \jacobiLength_1^{(\mutAlpha,\mutBeta)},
\ldots)$ and $\eFunLengthMatrix:=\diag(\eFunLength
_0, \eFunLength_1, \ldots)$ denote diagonal matrices with
entries $\jacobiLength_n^{(\mutAlpha,\mutBeta)}$ and $\eFunLength
_n$ defined as in Proposition~\ref{propinitial}. Then, for each ${k}
\in\{1,\ldots,\numPoints\}$, the coefficients in the representation
of the densities $g_{k}({y})$ and
$f_{k}({y})$ in (\ref{eqforwardDensityrepresentation}) and (\ref
{eqauxiliaryDensityrepresentation}) can be computed recursively as
%
\begin{eqnarray}
\vecAuxiliary_{k}&=& \vecForward_{{k}-1} \exp\bigl[ -
\eValMat(\diffTime_{k}- \diffTime_{{k}-1}) \bigr],
\label{eqforwardtoaux}
\\
\vecForward_{k}&=& \vecAuxiliary_{k}\eVecMatrix
\mathbf{G}^{{d}_{k}} (\bOne- \mathbf{G} )^{\numSamples_{k}- {d}_{k}}
\eVecMatrix^{-1},
\label{eqauxtoforward}
\end{eqnarray}
where $\eVecMatrix^{-1}$ is given by
%
\begin{eqnarray}
\eVecMatrix^{-1} &=& \jacobiLengthMatrix\eVecMatrix^T
\eFunLengthMatrix^{-1}. \label{eqeVecMatrixinv}
\end{eqnarray}
\end{theorem}

Combining Proposition~\ref{propinitial} and Theorem~\ref
{thmtransitionemission}, we obtain a dynamic programming algorithm
for calculating the coefficients $\vecForward_{k}$ and
$\vecAuxiliary_{k}$ in the representations for
$f_{k}$ and $g_{k}$
given in (\ref{eqforwardDensityrepresentation}) and (\ref
{eqauxiliaryDensityrepresentation}), respectively. The vectors and
matrices appearing in the above results are infinite dimensional. As in
previous works [\citet{Song2012,Steinrucken2013}] on the spectral\vadjust{\goodbreak}
representation of the transition density, when applying the above
results we truncate the infinite vectors and matrices by choosing
cutoffs for the dimensions. We provide more practical details in
Section~\ref{secperformance}.

Finally, the probability of observing the full data ${O}
_{[1\dvtx\numPoints]}$ can be computed using the following proposition:

%
\begin{proposition}\label{propfinalstep}
The probability ${\mathbb{P}}_\Theta\{ {O}_{[1\dvtx\numPoints]} \}
$ of observing the data ${O}_{[1\dvtx\numPoints]}$ given
the population genetic parameters $\Theta$ is
%
\begin{equation}
\label{eqfinalprob} {\mathbb{P}}_\Theta\{ {O}_{[1\dvtx\numPoints]} \} =
\frac
{\eFunLength_0 }{{B}_0(0)} \coeffForward_{\numPoints,0},
\end{equation}
where ${B}_0(0)$ is given by
\[
{B}_0(0) = \sum_{{m}=0}^\infty(-1)^{m}{w}_{0,{m}}
\frac{\Gamma({m}+\mutAlpha)}{\Gamma({m}+1)\Gamma(\mutAlpha)}.
\]
\end{proposition}

\section{Results}
\label{secresults}

In this section we perform parametric inference via the maximum
likelihood framework using a finite grid in the parameter space. We
first test the accuracy on simulated data and then apply it to analyze
an ancient DNA data set related to coat coloration in domesticated
horses [\citet{Ludwig2009}].

Since ancient DNA data are often collected from only those loci which
are segregating at the present time, in our empirical study we
condition on observing at least one copy of the derived allele at the
last sampling time $\diffTime_\numPoints$. In particular, the
likelihood of the parameters is given by $L(\Theta):=
{\mathbb{P}}_\Theta\{ {O}_{[1\dvtx\numPoints]} \mid
{d}_\numPoints> 0 \}$. We chose to maximize
this function on a grid, since the algorithm described in the previous
section can be parallelized, thus allowing to efficiently evaluate the
likelihood under given parameters for several data sets at once.




\subsection{Performance on simulated data}
\label{secsimulation}

We simulated data under a discrete-time Wright--Fisher model with
several values for the effective population size and selection
coefficients. We chose the mutation probabilities to be $\mutA= \mutB
= 10^{-6}$ and the number of years per generation to be five years.
These parameters are similar to those considered by previous works that
analyzed time series allelic samples from the ASIP and MC1R loci in
horses [\citet{Ludwig2009,Malaspinas2012}]. In our simulations,
5\% of the population carried the mutant allele when it first became
positively selected.
We sampled 40 individuals at each of 10 time points over the course of
32,000 years.

We investigated the performance of our maximum likelihood estimator in
various scenarios of selection. Here, we present the results for the
following four particular selection schemes:
\begin{longlist}[2.]
\item[1.] Genic selection, in which the selective fitness of the
heterozygote is the arithmetic mean of the fitness of the two
homozygotes, that is, ${{s}_{1}}= {s}/2$ and ${{s}_{2}}
= {s}$.
\item[2.] Heterozygote advantage selection, in which ${{s}_{1}}=
{s}$ and ${{s}_{2}}= 0$.
\item[3.] Recessive selection, in which ${{s}_{1}}= 0$,
${{s}_{2}}= {s}$.
\item[4.] Dominant selection, in which ${{s}_{1}}= {s}$,
${{s}_{2}}= {s}$.\vadjust{\goodbreak}
\end{longlist}
For each scenario, we considered ${s}\in\{0, 0.001,
0.0025,0.005,0.01\}$ and simulated 200 data sets for each value of $s$.

%
\begin{figure}

\includegraphics{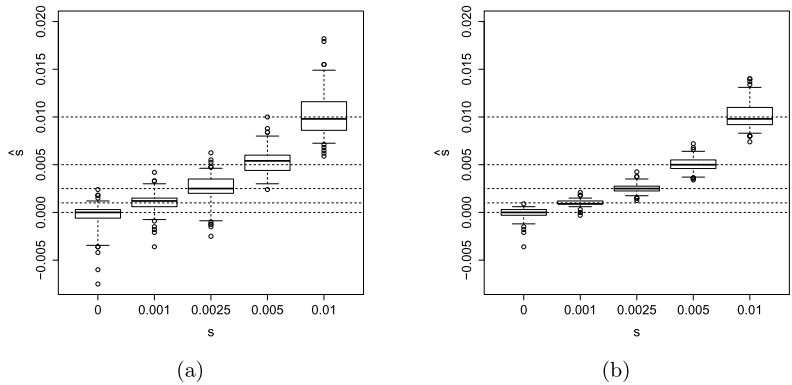}

\caption{Empirical distribution of the maximum likelihood estimates
for 200 data sets simulated under a model of genic selection, with
heterozygote fitness ${{s}_{1}}= {s}/2$ and derived allele
homozygote fitness ${{s}_{2}}= {s}$, for each of several
different values of selection strength ${s}$.
The dashed lines indicate the true values.
\textup{(a)} The effective population size $N_e$ is 2500
individuals.
\textup{(b)}~$N_e={}$10,000 individuals.}
\label{figgenict10}
\end{figure}

Figure~\ref{figgenict10} shows the performance of the maximum likelihood
estimator under a model of genic selection with an effective population
size of $N_e=2500$ and $N_e={}$10,000. It illustrates
empirical boxplots of the maximum likelihood estimates, where the tips
of the whiskers denote the 2.5\%-quantile and the \mbox{97.5\%-}quantile, and
the boxes represent the upper and lower quartile. As the figure shows,
our maximum likelihood estimates are unbiased.
The uncertainty of the estimate tends to increase with increasing
values of ${s}$, while the uncertainty decreases as the population
size increases, illustrating the fact that for larger population sizes,
selection acts more efficiently and is easier to detect. In the case of
$N_e={}$10,000, if the true selection coefficient is $0.0025$ or
more, all our maximum likelihood estimates are higher than the 97.5\%-quantile of the empirical distribution of the maximum likelihood
estimates for ${s}=0$. Hence, there is high power to reject
neutrality in these scenarios.

%
\begin{figure}

\includegraphics{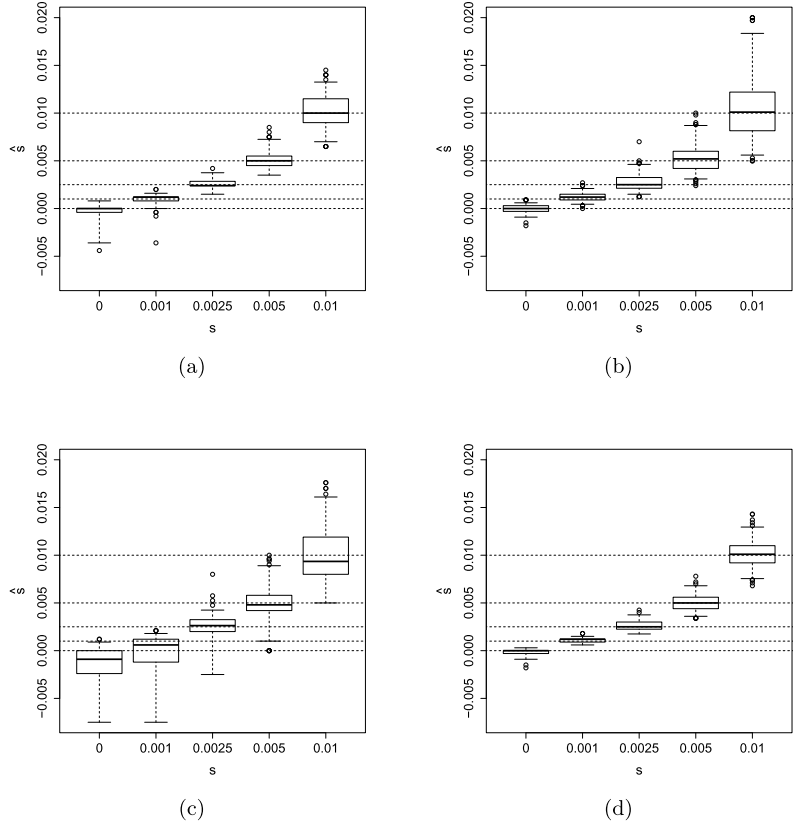}

\caption{Empirical distribution of the maximum likelihood estimates of~200 simulated data sets each under different modes of selection of
differing strength with $N_e={}$10,000.
The dashed lines indicate the true values of ${s}$.
\textup{(a)} Genic selection (${{s}_{1}}= {s}/2$,
${{s}_{2}}= {s}$) with only five sampling time points.
\textup{(b)} Heterozygote advantage model of selection
(${{s}_{1}}= {s}$, ${{s}_{2}}= 0$) with ten sampling
time points.
\textup{(c)}~Recessive selection (${{s}_{1}}= 0$,
${{s}_{2}}= {s}$) with ten sampling time points.
\textup{(d)} Dominant selection (${{s}_{1}}= {s}$,
${{s}_{2}}= {s}$) with ten sampling time points.}\vspace*{-3pt}\label{figadditionalsimulatedscenarios}
\end{figure}

The performance of our maximum likelihood estimator for several
additional selection schemes and parameter regimes can be found in
Figure~\ref{figadditionalsimulatedscenarios}, where we also
consider a
scenario with fewer sampling time points.
The figure shows that our maximum likelihood estimates are unbiased
across the different parameter ranges and scenarios. In general, the
low variance of the empirical distribution of the maximum likelihood
estimates shows that our method can be used to accurately infer the
selection parameters of interest in a wide range of scenarios.

\subsection{Analysis of ancient DNA data: Coat coloration in domesticated horses}\label{sechorses}

Ludwig et~al. (\citeyear{Ludwig2009}) extracted genotype data at several loci from
ancient horse DNA obtained from various sites in Eurasia. In
particular, they extracted\vadjust{\goodbreak} temporal allele frequency data at eight loci
that are known to play a role in coat color determination in
contemporary horses.
Only the locus encoding for the Agouti signaling peptide (ASIP) and the
locus for the melanocortin 1 receptor (MC1R) showed strong fluctuations
in the sample allele counts. Table~\ref{tabhorse} shows the time series
data for the ASIP and the MC1R loci in the curated form of the original
work [\citet{Ludwig2009}].

Using the method of \citet{Bollback2008} for the model of genic
selection (${{s}_{1}}= {s}/2$, ${{s}_{2}}= {s}$),
\citet{Ludwig2009} established that selection acted significantly
on only the ASIP and the MC1R loci. However, another recent analysis
[\citet{Malaspinas2012}] of the same data set considered the
model of recessive selection (${{s}_{1}}= 0$, ${{s}_{2}}=
{s}$) and did not find a significant signal of selection at the ASIP locus.

%
\begin{table}
\tabcolsep=0pt
\caption{The temporal allele frequency data sets for the ASIP and MC1R
loci associated with coat coloration in domesticated
horses~[\citet{Ludwig2009}, Figure~\textup{S.3}]. For each sampling time
${t}_{{k}}$ (given in years BCE), the table lists the
number ${d}_{{k}}$ of derived alleles among the
sampled $\numSamples_{{k}}$ alleles}\label{tabhorse}
\begin{tabular*}{\tablewidth}{@{\extracolsep{\fill}}@{}ld{5.0}d{5.0}d{4.0}d{4.0}d{4.0}d{4.0}@{$\!$}}
\hline
Time of sampling [${t}_{k}$] (BCE) & 20{,}000 & 13{,}100 &3700 & 2800 & 1100 & 500\\
\# of samples [$\numSamples_{k}$] & 10 & 22 & 20 & 20 & 36 &38\\[3pt]
ASIP (\# der. alleles) [${d}_{k}$] & 0 & 1 & 15& 12 & 15 & 18\\
MC1R (\# der. alleles) [${d}_{k}$] & 0 & 0 & 1& 6 & 13 & 24\\
\hline
\end{tabular*}
\end{table}

To investigate the dependence of the previous conclusions on the
assumed selection scheme, we applied our method to reanalyze the ASIP
and the MC1R data under a general selection scheme with arbitrary
selection coefficients ${{s}_{1}}$ and ${{s}_{2}}$.
We set the mutation probability to $\mutA= \mutB= 10^{-6}$ and the
average length of a generation to 5 years. Table~\ref{tabhorse} shows that
the derived allele is absent in both data sets at time 20,000 BCE.
Thus, we set the initial frequency of the derived allele as
$1/2N_e$, corresponding to the case where the selected allele
arises as a \emph{de novo} mutation at time ${t}_0$. We tried a
range of values for $N_e$ and ${t}_0$.

\begin{figure}

\includegraphics{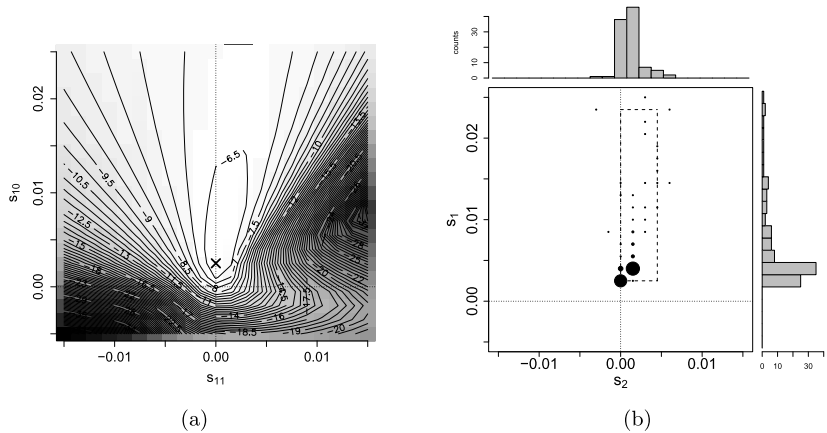}

\caption{Analysis of the ASIP locus.
\textup{(a)}~Empirical values of the likelihood $L(\Theta
)$ for temporal samples from the ASIP locus where the likelihood is
computed over a $21\times21$ grid.
The maximum is attained at $({{s}_{1}}, {{s}_{2}}) =
(0.0025, 0)$, indicated by the ``\textup{x}.''
\textup{(b)}~A joint density plot and marginal histograms of the
maximum likelihood estimates for 100 bootstrap resampled data sets of
the temporal data at the ASIP locus. The circles are centered on the
grid points at which the likelihood function is evaluated, and the
sizes of the circles indicate the proportion of maximum likelihood
estimates that occupy the same grid point. The marginal empirical 2.5\%
and 97.5\%-quantiles are $[0.0025,0.0235]$ for the heterozygote fitness
${{s}_{1}}$, and $[0,0.0045]$ for the derived allele homozygote
fitness ${{s}_{2}}$, as indicated by the dashed box.}\label{figrealASIP}\label{figsingleASIP}\label{figbootstrapASIP}
\end{figure}

\begin{figure}

\includegraphics{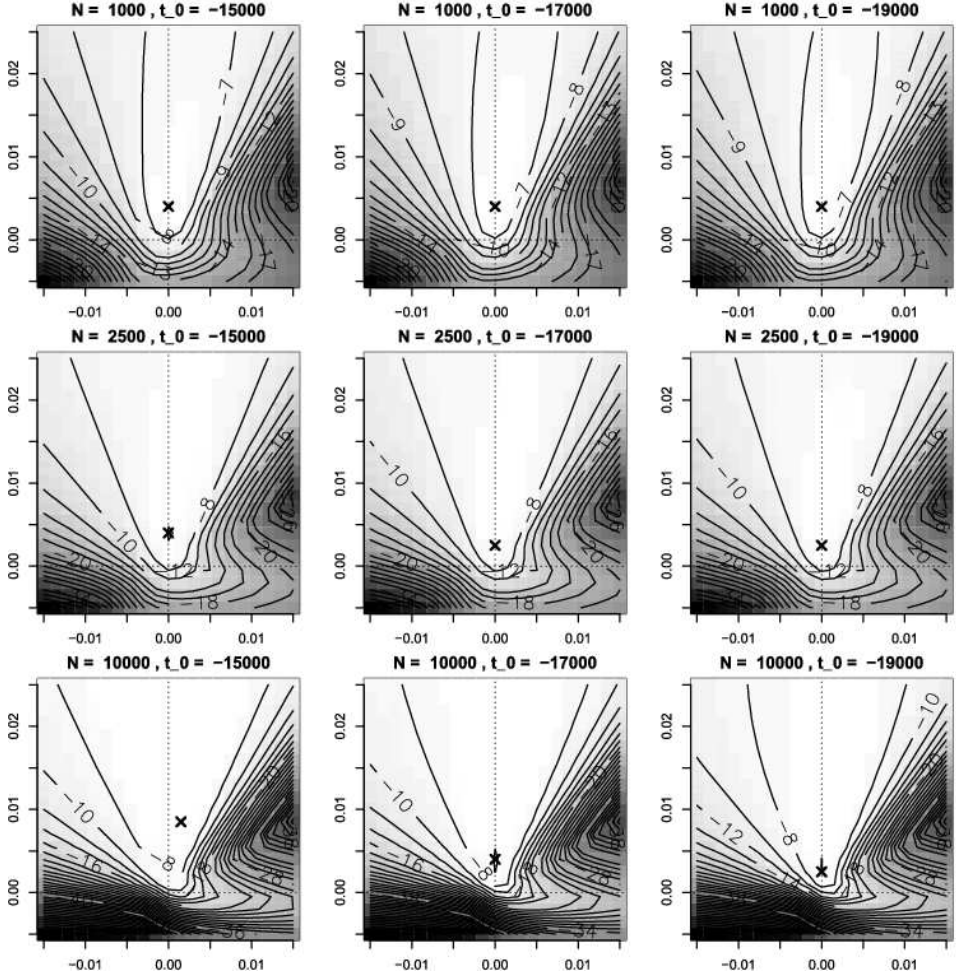}

\caption{Likelihood surfaces for the ASIP data set under various
combinations of $N_e\in{}$\{1000, 2500, 10,000\} and ${t}
_0 \in{}$\{15,000~BCE, 17,000~BCE, 19,000~BCE\}. The respective maxima are indicated by an ``\textup{x}.''}
\label{figmultipleASIP}
\end{figure}

Figure~\ref{figsingleASIP}(a) shows the likelihood surface for the temporal
allele frequency data from the ASIP locus, for $N_e= 2500$ and
${t}_0 ={}$17,000 BCE.
The empirical maximum of the likelihood surface is located at
$({{s}_{1}}, {{s}_{2}}) = (0.0025, 0)$, indicated by the
``\textup{x}'' in Figure~\ref{figsingleASIP}(a). This maximum suggests that a
selective scheme of heterozygote advantage best explains the data,
where both the ancestral and derived allele homozygotes are of equal
fitness, while the heterozygous genotype confers a selective advantage
over the homozygotes. To establish the significance of this finding, we
performed the following bootstrap procedure: we resampled the ASIP data
set 100 times to obtain subsampled data sets $ \{{O}
_{[1\dvtx\numPoints]}^{({j})} \}_{{j}=1}^{100}$.
For each bootstrapped data set $1 \leq{j}\leq100$, we
resampled $\numSamples_{k}^{({j})} = \numSamples
_{k}$ alleles at each time ${t}_{k}
^{({j})} = {t}_{k}$.
The number of derived alleles for data set ${j}$ was obtained
by binomial sampling from the empirical frequency of derived alleles in
the original ASIP data set, that is,
\[
{d}_{k}^{({j})} \sim{\xi} \biggl( \cdot; \numSamples_{k}^{({j})},
\frac
{{d}_{k}}{\numSamples_{k}} \biggr).
\]
We then reported the empirical maximum of the likelihood surface for
each of these resampled data sets.
Figure~\ref{figbootstrapASIP}(b) shows the empirical maximum likelihood
estimates and marginal histograms of the maxima for the 100 resampled
data sets. The marginal 2.5\% and 97.5\% quantiles of the empirical
distribution are $[0.0025,0.0235]$ for the heterozygote fitness
${{s}_{1}}$ and $[0,0.0045]$ for the derived allele homozygote
fitness ${{s}_{2}}$, thus providing further evidence that the
data are significantly better explained by a selection model where a
heterozygous individual is selectively advantageous over the homozygous
individuals.
As Figure~\ref{figmultipleASIP} shows, changing $N_e$ from 1000
to 10,000, or changing ${t}_0$ from 19,000 BCE to
15,000, BCE has only a minimal effect on the shape of the
likelihood surface and maximum likelihood estimate, again supporting
that a selective scheme of heterozygote advantage best explains the data.

%

%

\begin{figure}

\includegraphics{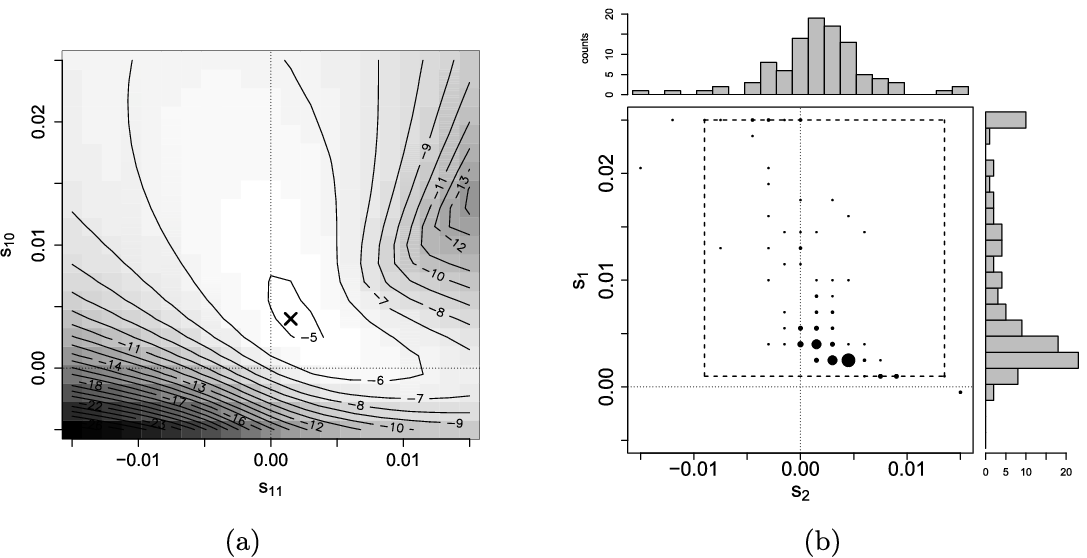}

\caption{Analysis of the MC1R locus using the parameters $N_e
=2500$ and ${t}_0 = 7000$ BCE.
\textup{(a)}~Likelihood surface for the MC1R locus. The maximum
likelihood estimate is
at $({{s}_{1}},{{s}_{2}}) = (0.004,0.0015)$ and is
indicated by the ``\textup{x}.''
\textup{(b)}~A joint density plot and marginal histograms of the maximum
likelihood estimates for 100 bootstrap resampled data sets obtained
from the
MC1R data as described in Section~\protect\ref{sechorses}. The
marginal 2.5\% and 97.5\%-quantiles
are $[0.001, 0.025]$ for the heterozygote fitness $s_1$ and $[-0.009,
0.0135]$ for the
derived allele homozygote fitness $s_2$, as indicated by the dashed
box.}\label{figmc1r}\vspace*{-3pt}
\end{figure}

\begin{figure}

\includegraphics{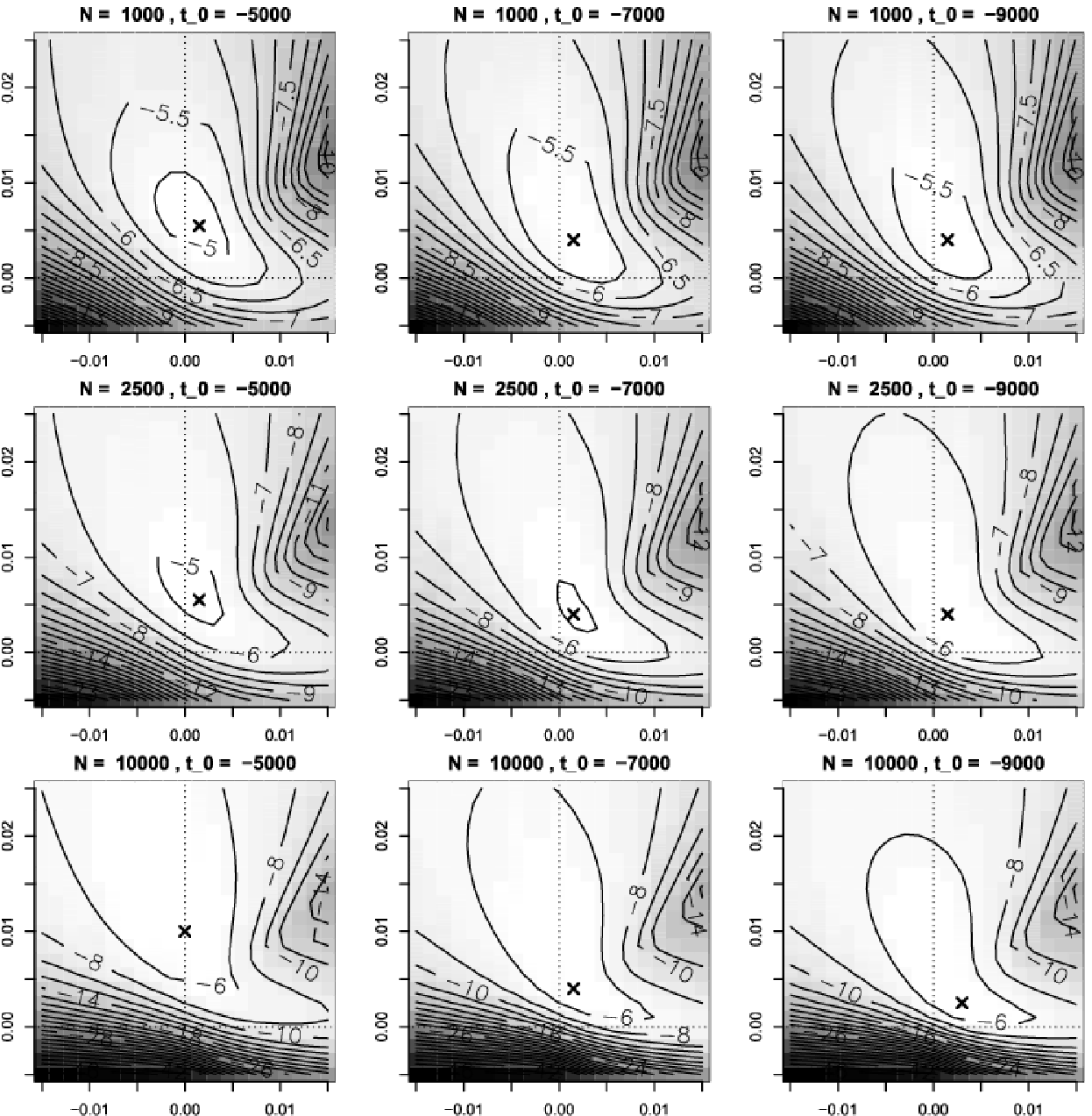}

\caption{Likelihood surfaces for the MC1R data set under various
combinations of $N_e\in{}$\{1000, 2500, 10,000\} and ${t}
_0 \in{}$\{5000~BCE, 7000~BCE, 9000~BCE\}. The respective maxima are indicated by an ``\textup{x}.''}
\label{figmultipleMC1R}
\end{figure}

A similar analysis of the MC1R locus can be found in Figures~\ref
{figmc1r} and~\ref{figmultipleMC1R}. For this data set, the maximum
of the likelihood surface is attained at $({{s}_{1}},{{s}_{2}}) =
(0.004,0.0015)$, and the empirical marginal 2.5\%
and 97.5\%-quantiles are $[0.001, 0.025]$ for the heterozygote fitness
and $[-0.009, 0.0135]$ for the derived allele homozygote fitness.
Together with the results shown in Figure~\ref{figmultipleMC1R},
this suggests that the data at the MC1R locus is also best explained by
a selection model of heterozygote advantage. However, although the
marginal quantiles for the homozygote fitness cover ${{s}_{2}}=
0$, they are rather far apart, so the evidence of heterozygote
advantage for the MC1R locus is weaker than that for the ASIP locus.

%

%

\subsection{Computational performance}\label{secperformance}

The running time of our algorithm for computing the likelihood of a
given set of population-scaled\vspace*{2pt} parameters is dependent on the
dimensions of the truncation $\widetilde{\mathbf{M}}$ of the
infinite matrix $\mathbf{M}$ given in (\ref{eqoperatormatrix}). In
particular, the time complexity of computing a
single likelihood is the cost of computing the eigenvalues and
eigenvectors of $\widetilde{\mathbf{M}}$ plus the cost of
computing the coefficients $\vecForward_{k}$ in Theorem~\ref{thmtransitionemission}, where $k \in\{1, \ldots, K\}$.
To compute the eigenvalues and eigenvectors of $\widetilde
{\mathbf{M}}$ to high precision, we first used LAPACK\footnote
{Available from \url{http://www.netlib.org/lapack/}.} to compute them
to double precision, and then refine them by using inverse iteration
[\citet{Press1992}, Chapter~11.8]. Each step of the inverse
iteration involves solving a linear system with matrix $\widetilde
{\mathbf{M}} - \mu\mathbf{I}$, where $\mu$ is an estimate for
an eigenvalue of $\widetilde{\mathbf{M}}$. Since this matrix has
bandwidth at most 9, this linear system can be solved in $O(D)$ time,
where $D$ is the dimension of $\widetilde{\mathbf{M}}$.
By using the repeated squaring algorithm for taking powers of the
matrices $\mathbf{G}$ and $\bOne- \mathbf{G}$ and exploiting the
fact that $\mathbf{G}$ and $\bOne- \mathbf{G}$ are tridiagonal
matrices, each coefficient $\vecForward_{k}$ can be computed
in $O(D^2 + D \min(D, n_k)^2 \log n_k)$ time, where the first $O(D^2)$
term comes from the matrix-vector multiplications in (\ref{eqauxtoforward}).


For the analysis of the ASIP and MC1R data sets reported in
Figures~\ref{figrealASIP} and~\ref{figmc1r}, we approximated the
eigenvalues and eigenvectors of $\mathbf{M}$ defined in (\ref
{eqoperatormatrix}) using a $600\times600$ submatrix. Furthermore, we
used the first $590$ terms in~(\ref{eqdefefun}) to approximate the
eigenfunctions, and the dimensions of the vectors of coefficients
in~(\ref{eqvecForward}) and (\ref{eqvecAuxiliary}) were set to
$580$. We empirically verified that these cutoffs produced a stable
approximation of the likelihood. Using these values, the computation
time for a single point of the grid in Figure~\ref{figsingleASIP}(a) was
approximately 95 seconds. We adjusted the cutoffs appropriately for the
other analyses reported in Section~\ref{secresults}.

\section{Discussion}\label{secdiscussion}

In this paper we have developed a novel, efficient spectral algorithm
to analyze time series allele frequency data under a general diploid
selection model.
We have demonstrated that our method can be used to accurately estimate
selection parameters on simulated data.

We have also applied our method to investigate loci involved in horse
coat coloration. Our inferred selection\vadjust{\goodbreak} coefficients show that the data
are best explained by a heterozygote advantage model of balancing selection.
As mentioned earlier, \citet{Ludwig2009} provided evidence for
slightly positive selection at the ASIP locus, assuming a model of
genic selection (where ${{s}_{1}}= {{s}_{2}}/2$). More
precisely, they obtained a point estimate of ${{s}_{2}}= 0.0007$
and a 95\% confidence interval of $[0.0001,0.0015]$. However, using a
model of selection where the derived allele homozygote is recessive
(i.e., ${{s}_{1}}=0$), a subsequent reanalysis [\citet
{Malaspinas2012}] of the same data found that ${{s}_{2}}$ has a
point estimate of $-$0.001 with a 95\% confidence interval of
$[-0.02,0.051]$,\vadjust{\goodbreak} thus not rejecting neutrality at the ASIP locus.
In our work, we have allowed our method to explore the two-dimensional
parameter space of general diploid selection models and presented
evidence for a selection mode where heterozygous individuals are
advantageous over homozygous individuals.
It is possible that previous analyses have only been able to infer very
weak selection acting at the ASIP locus because they have restricted
the model of selection to certain one-dimensional models. Indeed, if we
restrict our analysis to a model of genic selection, we get results
similar to those reported by \citet{Ludwig2009}.
Our analysis does not conclusively prove that individuals that were
heterozygous at the ASIP locus had a constant evolutionary advantage
since 17,000 BCE, because we have ignored the interaction of
selection and demographic history, epistatic interactions between loci,
time-varying models of selection and other factors.
However, our results suggest the possibility that some mode of
heterozygote advantage balancing selection has maintained polymorphism
at the ASIP locus that is involved in horse coat coloration.

Although we have focused on time series samples taken at a biallelic
locus, the mathematical framework presented here could be readily
extended to handle an arbitrary number of alleles using the spectral
representation derived by \citet{Steinrucken2013}.
Further, changes in the population size and selection coefficients
could be modeled by suitably combining the spectral representations for
different population genetic parameters at the change points. It is
also possible to extend the method to multiple populations and to
incorporate samples taken from extinct ancestral populations. In light
of emerging ancient DNA sequence data for ancient hominids [\citet
{Green2010,Reich2010}], such temporal sequence data and inference
methods present novel opportunities to gain insight into adaptation in
humans. For a more adequate modeling of biologically relevant
scenarios, it is also necessary to incorporate the exchange of migrants
into the model [\citet{Gutenkunst2009,Lukic2011}] and extend the
framework to incorporate variation at linked loci. By taking advantage
of genetic hitchhiking at closely linked sites during the course of
selective sweeps, one might be able to further improve the inference of
selection coefficients.

\section*{Acknowledgments}
We thank Rasmus Nielsen, Joshua Schraiber and\break Montgomery Slatkin for
helpful comments and discussions. We also thank Karen Kafadar and two
anonymous referees for suggestions that improved the exposition of this
paper. Moreover, we thank Richard J. Mathar [\citet{Mathar2009}]
for making his source code available to us.

\begin{supplement}[id=supplement]
\stitle{A novel spectral method for inferring general diploid selection from time series genetic data}
\slink[doi]{10.1214/14-AOAS764SUPP} 
\sdatatype{.pdf}
\sfilename{aoas764\_supp.pdf}
\sdescription{We provide proofs of the results stated in Section~\ref{secmodel}. The modified Jacobi polynomials appearing in this paper
are defined and some of their key properties are listed. Also, the
coefficients in the definition of the matrix $\mathbf{M}$ in
equation~(\ref{eqoperatormatrix}) are provided. Last, we describe
some alternate density functions for the allele frequency at the time
when selection arises.}
\end{supplement}

%

\printaddresses

\end{document}